\documentclass[journal]{IEEEtran}
\usepackage{xspace}
\makeatletter
\DeclareRobustCommand\onedot{\futurelet\@let@token\@onedot}
\def\@onedot{\ifx\@let@token.\else.\null\fi\xspace}

\def\ie{i.e\onedot} 
 
\def\etc{etc\onedot}

\makeatother

\ifCLASSOPTIONcompsoc
    \usepackage[caption=false, font=normalsize, labelfont=sf, textfont=sf]{subfig}
\else
\usepackage[caption=false, font=footnotesize]{subfig}
\fi
\usepackage[hidelinks]{hyperref}
\usepackage{cite}
\usepackage{multirow}
\usepackage{diagbox}
\usepackage{amsmath,amssymb,amsfonts}
\usepackage{graphicx}
\usepackage{algorithm}
\usepackage{algpseudocode}
\usepackage[table,xcdraw]{xcolor}
\usepackage{bm}
\usepackage[inline]{enumitem}

\newtheorem{theorem}{Theorem}
\newtheorem{definition}{Definition}
\newtheorem{lemma}{Lemma}

\usepackage{commath}
\usepackage[capitalize]{cleveref}
\Crefname{section}{Section}{Sections}
\Crefname{table}{Table}{Tables}
\Crefname{algorithm}{Algorithm}{Algorithms}
\Crefname{theorem}{Theorem}{Theorems}
\Crefname{lemma}{Lemma}{Lemmas}
\Crefname{definition}{Definition}{Definitions}
\Crefname{proposition}{Proposition}{Propositions}

\newcommand{\JQ}[1]{#1}

\newcommand\myleq[1]{\mathrel{\overset{\makebox[0pt]{\mbox{\normalfont\tiny\sffamily (#1)}}}{\leq}}}

\begin{document}

\title{Dependence Analysis and Structured Construction for Batched Sparse Code}

\author{
        \IEEEauthorblockN{Jiaxin~Qing,
        Xiaohong~Cai,
        Yijun~Fan,
        Mingyang~Zhu,
        and~Raymond~W.~Yeung,~\IEEEmembership{Fellow,~IEEE}}%
\thanks{%
    J.~Qing, X.~Cai, and Y.~Fan are with the Department of Information Engineering, The Chinese University of Hong Kong, Hong Kong SAR (emails: \{\href{mailto:jqing@ie.cuhk.edu.hk}{jqing}, \href{mailto:cx021@ie.cuhk.edu.hk}{cx021}, \href{mailto:fy022@ie.cuhk.edu.hk}{fy022}\}@ie.cuhk.edu.hk).}%

\thanks{M.~Zhu
is with the Institute of Network Coding, The Chinese University of Hong Kong (email: \href{mailto:mingyangzhu@cuhk.edu.hk}{mingyangzhu@cuhk.edu.hk})}

\thanks{R.~W.~Yeung is with the Department of Information Engineering, The Chinese University of Hong Kong, Hong Kong SAR. R.~W.~Yeung is also with the Institute of Network Coding, The Chinese University of Hong Kong, and he is also a Principal Investigator of the Centre for Perceptual and Interactive Intelligence (CPII) Limited (email: \href{mailto:whyeung@ie.cuhk.edu.hk}{whyeung@ie.cuhk.edu.hk}). The work of R. W. Yeung was supported in part by a fellowship award from the Research Grants Council of the Hong Kong Special Administrative Region, China under Grant CUHK SRFS2223-4S03.}

}

\maketitle

\begin{abstract}
In coding theory, codes are usually designed with a certain level of randomness to facilitate analysis and accommodate different channel conditions. However, the resulting random code constructed can be suboptimal in practical implementations. Represented by a bipartite graph, the Batched Sparse Code (BATS Code) is a randomly constructed erasure code that utilizes network coding to achieve near-optimal performance in wireless multi-hop networks. 
In the performance analysis in the previous research, it is implicitly assumed that the coded batches in the BATS code are independent. This assumption holds only asymptotically when the number of input symbols is infinite, but it does not generally hold in a practical setting where the number of input symbols is finite, especially when the code is constructed randomly. 
We show that dependence among the batches significantly degrades the code's performance. \JQ{In order to control the batch dependence through graphical design,} we propose constructing the BATS code in a structured manner.  A hardware-friendly structured BATS code called the Cyclic-Shift BATS (CS-BATS) code is proposed, which constructs the code from a small base graph using light-weight cyclic-shift operations. We demonstrate that when the base graph is properly designed, a higher decoding rate and a smaller complexity can be achieved compared with the random BATS code. 

\end{abstract}

\section{Introduction}

\IEEEPARstart{T}{he} sixth-generation communication (6G) is envisioned to be reliable and intelligent, providing seamless connectivity for global computing and broadband coverage. It will be the key infrastructure that supports an even higher density of connections from a wider variety of devices than 5G, such as mobile phones, vehicles, wireless sensors, and other edge devices, creating a massive wireless network where different nodes are connected and communicating through each other~\cite{6G-1,6G-2}. The wireless multi-hop network is a commonly used model to study data transmission in wireless mesh networks and empowers a wide range of applications, such as integrated ground-air-space networks~\cite{space-air-ground},  smart sensing~\cite{wsn}, autonomous driving~\cite{autodrive, auto_drive}, internet-of-things~\cite{iot} and integrated access-backhaul networks (IAB)~\cite{iab}, \etc, which is crucial for providing a seamless, stable and intelligent communication experience for users. 

\begin{figure}
\centering
\includegraphics[width=.9\linewidth]{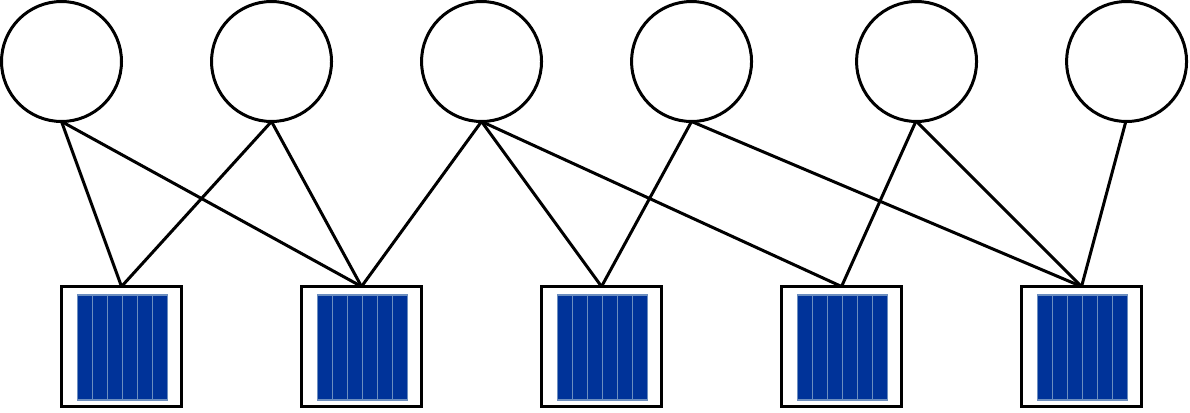}
\caption{\textbf{Graphical Representation of the BATS Code.} The input symbols are represented by circles (variable nodes). The batches are represented by squares (check nodes), which consist of several coded symbols. }
\label{fig:small_cycle}
\end{figure}

However, data loss is inevitable in wireless communication because of certain physical phenomena, such as refraction, diffraction, and multipath reflection, as electromagnetic waves propagate through the air. 
From the upper layer protocols' perspective, data are represented as packets. An erasure channel can model the data transmission, where a packet is either lost or well-received.  In addition to the loss induced in the physical layer, packets can also be lost due to channel congestion and competition or unreliable connections due to the high mobility of devices~\cite{loss-1,loss-2}. 
As a result, packet loss accumulates exponentially fast as data traverse through the multi-hop network, which can easily exceed the threshold that TCP can handle with packet retransmissions after a few hops.\footnote{TCP is based on the assumption that the loss is below an outrage probability. 
This assumption is always true when the optical fiber or the twisted pair cable is used~\cite{tcp-hop,tcp-control}.} Even though some techniques exist to constrain the link-to-link packet loss under a certain threshold, for example, Adaptive Modulation and Coding (AMC), they rely on using either a higher transmission power or a lower data rate~\cite{amc,tcp-control,tcp-prob}, which makes reliable and high-throughput communication through a wireless multi-hop network impractical. 
According to \cite{mutihop_drops}, the throughput of a 20~Mbps single-hop network drops to around 1~Mbps when the number of hops increases to 8 using IEEE802.11a.

The Batched Sparse Code (BATS Code) solves this ``multi-hop curse'' with a network-coded fountain~\cite{net-fountain}, where the intermediate nodes perform coding on the received packets rather than simple forwarding. The end-to-end packet loss asymptotically converges to the single-hop packet loss by employing network coding. Using the BATS code, the network capacity can be approached for unicast networks and certain multicast networks under different network topologies~\cite{bats-tit}. 

However, when analyzing the performance~\cite{bats-tit,bats-tree,bats-finite}, an implicit assumption is made that check nodes are independent. But when the number of input symbols (variable nodes) and the number of coded symbols (check nodes) are moderately small, check nodes can be highly dependent. 
In this paper, we will show that the mutual independence of check nodes is a sufficient condition that achieves an upper bound of the decoding rate; the decoding rate decreases as the dependence strength increases. The dependence among check nodes is not considered in the random BATS code construction, where check nodes are randomly connected to variable nodes according to a designed degree distribution.

\begin{figure}
\centering
\includegraphics[width=\linewidth]{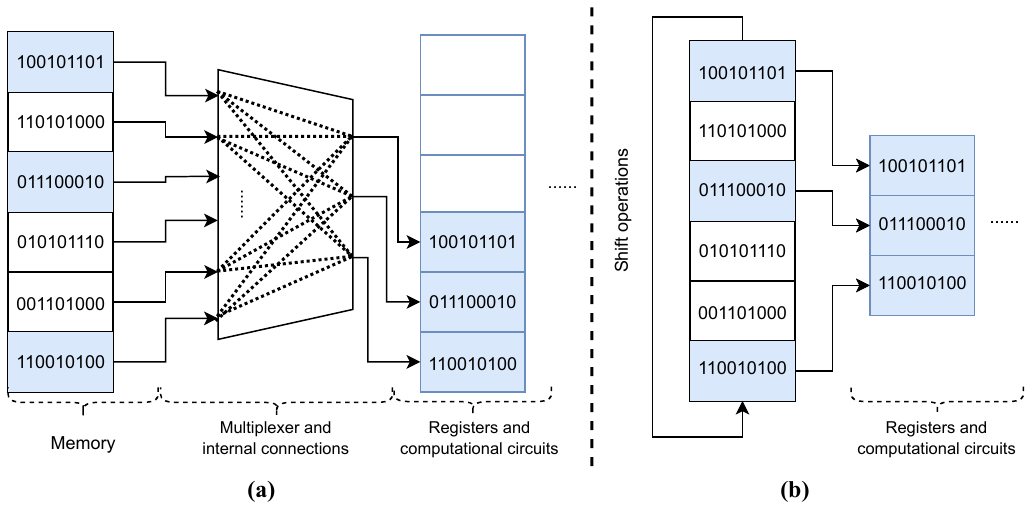}
\caption{\textbf{Random code construction complicates circuit routing in hardware implementations (A high-level illustration).} Comparision of the circuits between (a) randomly selecting input symbols for encoding and (b) selecting input symbols in a structured way using shift operations. Random selection leads to a fully connected circuit implementation in hardware.}
\label{fig:circuit_complexity}
\end{figure}

Additionally, in practical adaptations of the BATS code, a hardware implementation, for example, Field Programmable Gate Array (FPGA) or Application Specific Integrated Circuit (ASIC), is usually required to match the increasing demand for throughput and power-consumption efficiency~\cite{fpga-power,fpga-power2}. However, the traditional random construction leads to high complexity in circuit routing and buffer allocations in hardware implementations~\cite{ldpc-decoder-survey, ldpc-survey}. An example of added complexity due to random code construction is given in \cref{fig:circuit_complexity}a. To randomly select input symbols for encoding in hardware, a fully connected circuitry is needed to route all possible combinations of the input symbols from the memory to the register buffers for further computations. Furthermore, the number of symbols to select (\ie, degree) is usually determined on the fly according to a probability distribution. This means the register buffers must be sizeable enough to accommodate the maximum degree, even though its occurrence may be low. 

To address these issues, we propose constructing the BATS code in a more structured way so that better performance and lower complexity can be achieved, preserving the desired properties of a BATS code simultaneously.
We summarize our contributions as follows. 
\begin{enumerate}
\item We show that the check node dependence degrades the decoding rate\footnote{The decoding rate is defined as the number of decoded symbols divided by the total number of source symbols when the decoding stops.}, and independence is a sufficient condition to achieve the decoding rate upper bound. 
\item \JQ{We propose a new class of BATS codes called the structured BATS, exploiting a new design dimension and reducing node dependence in the construction. In particular, we introduce a hardware-friendly protograph-based Cyclic-Shift construction method (CS-BATS).} 
\item From exhaustive numerical simulations, we show that CS-BATS is superior not only in implementation complexity but also in decoding rate when the code is well-structured. 
\end{enumerate}

The remainder of this paper is organized as follows. \cref{sec:related-work} and \cref{sec:preliminary} present the related works and preliminaries.
\cref{sec:small-cycles} shows that check node dependence degrades the decoding rate of the BATS code from two perspectives: conditional probability and correlation. 
The Cyclic-Shift BATS code is introduced in \cref{sec:cs-bats}.
Simulations are performed in \cref{sec:cs-bats} to compare the performance and complexity of the CS-BATS and the random BATS. \cref{sec:layer-dec} presents a reduced complexity decoding scheme based on the CS-BATS code for hardware implementations. Finally, our conclusions are offered in \cref{sec:conclusion}.

\section{Related Work}
\label{sec:related-work}
\JQ{The BATS code retains the defining characteristics of the fountain code, including its rateless nature and the low complexity of its encoding and decoding processes.
When compared with traditional random linear network coding methods~\cite{netcod1, netcod2, netcod3}, the BATS code offers advantages such as lower complexity in encoding and decoding, reduced overhead for the coefficient vector, and decreased caching requirements at the intermediate nodes. Furthermore, compared with other low-complexity random linear network coding techniques such as the Gamma code~\cite{netcod4, netcod5}, EC code~\cite{netcod6}, and L-chunked code~\cite{netcod5}, the BATS code consistently delivers a higher transmission rate and can produce an unlimited number of batches. 
The utilization of the BATS code in various network communication settings has been examined in the studies in \cite{app1, app2, app3}. }
More discussion on the related works of the BATS code and their comparison could be found in \cite{bats-tit, bats-finite}.

The BATS code can be described and decoded graphically according to the Tanner graph~\cite{tanner-graph}, similar to the Low-Density Parity-Check (LDPC) code~\cite{ldpc,ldpc-mackay,ldpc-mac-tit}. 
Motivated by the protograph LDPC~\cite{tanner-qcldpc,ldpc-qc-pascal,ldpc-survey, qcldpc-girth1,qcldpc-girth2,qcldpc-girth3}, \cite{bats-rl} explored the design of a structured BATS code to increase the decoding rate when the number of input symbols is small. The authors in \cite{bats-rl} used a reinforcement learning approach to explore the graphical space, trying to find graphs that give a higher decoding rate and coding efficiency than the randomly constructed BATS code. However, this method fails to preserve the rateless property of the BATS code, and the deep learning models add extra overheads for practical implementations. Nonetheless, it shows that random construction is suboptimal regarding the decoding rate and complexity. 

Before this work, there were mainly three ways to analyze the performance of BATS codes: differential equation analysis~\cite{bats-tit}, tree analysis~\cite{bats-tree}, and finite-length analysis~\cite{bats-finite}. Differential equation analysis and tree analysis consider the asymptotic decoding rate as the number of input symbols goes to infinity. The finite-length analysis evaluates the decoding rate for a given number of batches, which is a more practical analysis relaxing the asymptotic assumption in the previous work. However, all these analyses implicitly assume that check nodes are decoded independently. 
For example, 
in the finite-length analysis, $p_{t,s}$, the probability that a batch is decodable for the first time at time $t$ and has degree $s$, is used to derive a recursive formula for calculating the decoding stopping time. 
In formulating $p_{t,s}$, a hypergeometric distribution is used~\cite{bats-finite}, which implicitly assumes check nodes are decoded independently. However, the decodable probability of check nodes can be highly dependent on each other when \JQ{the number of input symbols ($K$) is small}, especially when the Tanner graph is constructed randomly.  

\section{BATS Code Preliminary}
\label{sec:preliminary}
The BATS code is a matrix-generalized network-coded fountain code~\cite{fountain} that works on multi-hop networks with erasure channels. It enables an operation called ``recoding'' at the intermediate nodes, which performs random linear network coding (RLNC)~\cite{rlnc} on the received packets. The design of the recoding operation is called the inner code. 
Careful design of the inner code increases the network throughput in different scenarios~\cite{hoover1, hoover2, hoover3}.
Besides the inner code, the BATS code consists of an outer code that performs encoding and decoding at the source and destination nodes. In this work, we mainly study the outer code, and the recoding is included in this section for the completeness of the discussion. 

\subsection{Encoding}
As a matrix generation of the fountain code, the outer code generates encoded packets as ``batches'' comprising several coded packets. To generate a batch, we first need to randomly select $dg$ symbols from a total of $K$ input symbols, where $dg$ is a degree sampled from an optimized probability distribution $\Psi$. Each symbol is a vector of $pk$ elements from the Galois field $GF(q)$. Then, we take $M$ different linear combinations of the selected symbols to generate a batch of size $M$. This process can be described by the following linear system for generating the $i$-th batch,
\begin{equation}
    \bm{X}_i = \bm{B}_i\bm{G}_i,
    \label{eq:encoding}
\end{equation}
where $\bm{B}_i\in\mathbb{F}_q^{pk\times dg}$ is comprised of $dg$ selected symbols, and $\bm{G}_i\in\mathbb{F}_q^{dg\times M}$ describes the linear combinations to take, resulting in a matrix $\bm{X}_i\in\mathbb{F}_q^{pk\times M}$ representing $M$ coded symbols. The resulting code can be represented by a Tanner graph as in \cref{fig:small_cycle}.

\subsection{Recoding}
The recoding takes random linear combinations of the received packets from the same batch and generates a new batch of size $M$. It can be described by the linear system
\begin{equation}
    \bm{Y}_i = \bm{X}_i\bm{H}_i, \label{eq:bats-recode}
\end{equation}
where $\bm{Y}_i\in\mathbb{F}_q^{pk\times M}$ is the recoded batch, $\bm{X}_i\in\mathbb{F}_q^{pk\times m}$ consists of the received packets, $\bm{H}_i\in\mathbb{F}_q^{m\times M}$ is called the transfer matrix and $m$ is the number of received packets. If no packet is lost during the transmission, $m$ will equal $M$. If $m$ is smaller than $M$, which means packets are lost during the transmission, the rank of the batch will decrease even though the recoding generates $(M-m)$ packets. 
Therefore, we can use a rank distribution to model the end-to-end channel condition from the source node to the destination node. Specifically, the rank distribution is written as $h = (h_0,\ldots,h_j,\ldots,h_M)$, where $h_j$ represents the probability of receiving a batch with a rank of $j$, and $j\in[0,M]$.

\subsection{Decoding}
The BATS code is decoded at a destination node where the $K$ input symbols need to be recovered. 
Consider the system of linear equations received at a destination node,
\begin{equation}
    \bm{Y}_i=\bm{B}_i\bm{G}_i\bm{H}_i, \label{eq:bats-decode}
\end{equation}
where $\bm{Y}_i\in\mathbb{F}_q^{pk\times m}$ consists of the received packets, $\bm{B}_i\in\mathbb{F}_q^{pk\times dg_i}$ consists of selected input symbols and $\bm{G}_i\bm{H}_i\in\mathbb{F}_q^{dg_i\times m}$ is the product of the generator matrix and the transfer matrix. In (\ref{eq:bats-decode}), $\bm{Y}_i$ and $\bm{G}_i\bm{H}_i$ are known from the received information while $\bm{B}_i$ comprises the unknown symbols to be solved. 
For each batch, we need to solve this linear system and recursively substitute the already decoded symbols to other batches according to the underlying Tanner graph.  
This recursive linear-equation-solving and substituting decoding process is called the belief propagation (BP) decoding for the BATS code.
A batch is decodable if and only if the rank of the matrix formed by the received packets (the rank of the received packets for brevity) is equal to the degree of this batch, namely $\textrm{rank}(\bm{G}_i\bm{H}_i)= dg_i$. 


However, the performance of BP decoding is not satisfactory when the number of input symbols is small. Thus, a compromise between BP and the Gaussian elimination decoding, called inactivation decoding~\cite{bats-tit}, is usually used for recovering more input symbols at the expense of higher computational complexity. 
The induced complexity of inactivation decoding increases as the number of inactivation symbols increases. In the extreme case that all the unsolved variables are inactivated, it becomes Gaussian elimination decoding.  

\subsection{Degree Distribution Optimization}
\label{sec:deg_opt}
Given a rank distribution $h = (h_0,\ldots,h_M)$ at a destination node, the most important step in designing a good BATS code for that destination node is to find the asymptotically optimal degree distribution $\Psi$, which can be obtained by solving the following optimization problem,  
\begin{subequations} \label{eq:bats-opt}
\begin{align}
\max_{\Psi} \quad   & \theta \label{eq:bats-opt-obj}\\
\textrm{s.t.} \quad & \Omega(x,\Psi,h) + \theta\ln(1-x)\geq 0, \ \forall\ 0\leq x\leq\eta\label{eq:bats-opt-nec} \\ 
                    & \sum^K_{d=1}\Psi_d=1, \ \Psi_d \geq 0, \quad d= 1, ..., K\label{eq:bats-opt-sum}
\end{align}
\end{subequations}
where $\theta$ is the achievable rate. The definitions of $\Omega$ and $\zeta_r^k$ can be found in \cite{bats-tit}. 
From the tree analysis~\cite{bats-tree} and asymptotic analysis~\cite{bats-tit} of the BATS code, (\ref{eq:bats-opt-nec}) gives a sufficient condition for decoding up to $\eta K$ input symbols with probability at least $1-\exp({-cK})$.

\section{Decoding Rate and Dependence}
\label{sec:small-cycles}

The decoding rate is defined as the portion of variable nodes decoded by the end of the decoding, which can be analyzed from the decodable probability of an arbitrary variable node. 
\JQ{Notably, the decoding rate is not affected by the order in which the batches are decoded~\cite{bats-tit}.}
The decodable probability is defined as starting from the initial state, the probability of being decoded when the decoding stops.
In this section, we investigate how this probability would change if the dependence relation among its neighboring check node changes. 

\JQ{%
In this section, we associate each node (variable node and check node) with a Bernoulli random variable as an indicator of its decodability. 
Firstly, we analyze the expectation of a variable node under a simplified model, where we investigate how this expectation would change with the correlation between two neighboring check nodes.
Then, a lower bound and an upper bound will be derived for the decodable probability of a variable node in terms of the decodable probability of its neighboring check nodes under a general model.}

\subsection{Correlation Decreases Decodable Expectation}

\begin{figure}[t]
    \centering
    \subfloat[$\rho_{C_{k1},C_{k2}}=1$]{\includegraphics[width=0.4\linewidth]{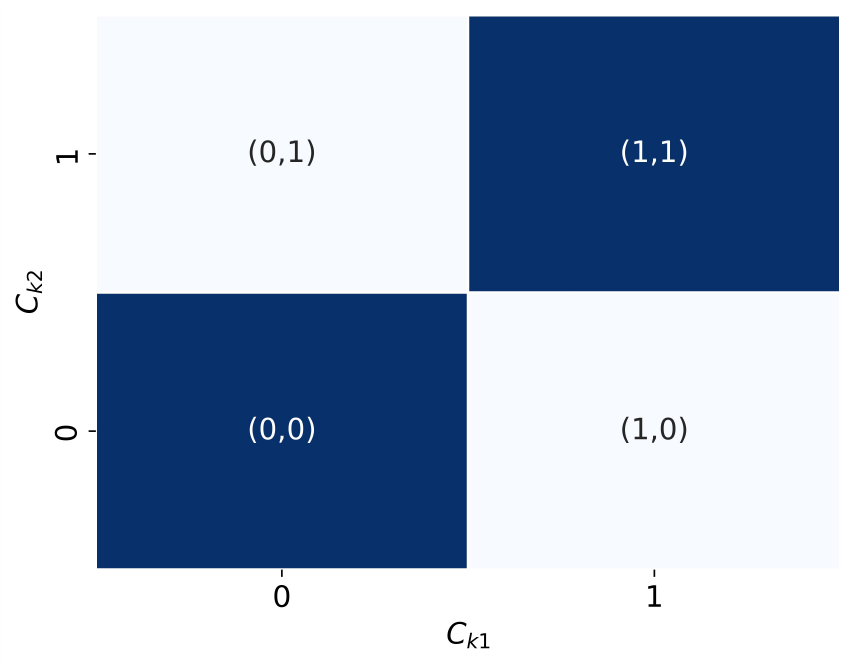}\label{fig:cor_plots_a}}
    \subfloat[$\rho_{C_{k1},C_{k2}}=0$]{\includegraphics[width=0.4\linewidth]{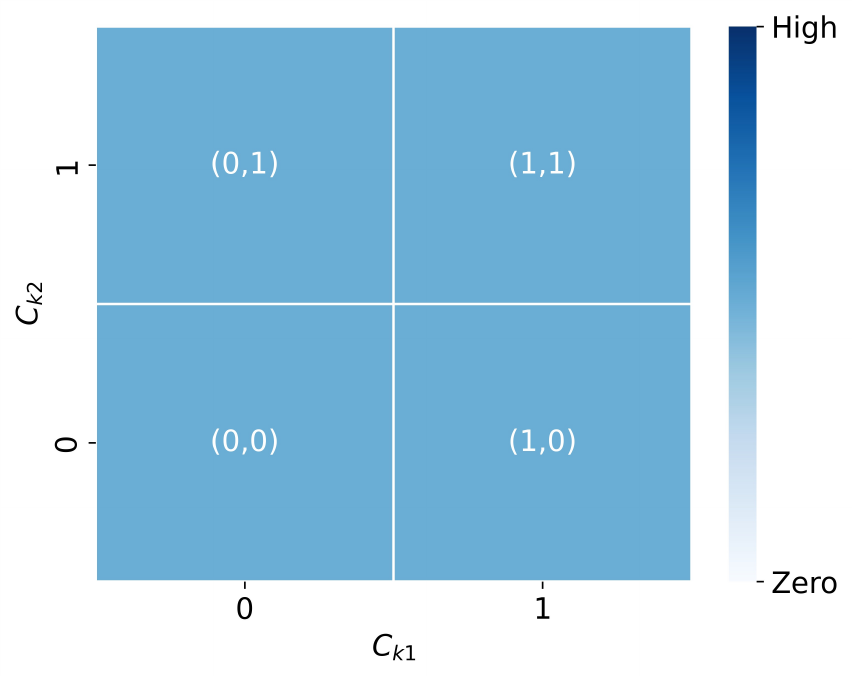}\label{fig:cor_plots_b}}
    
    \subfloat[$\rho_{C_{k1},C_{k2}}>0$]{\includegraphics[width=0.4\linewidth]{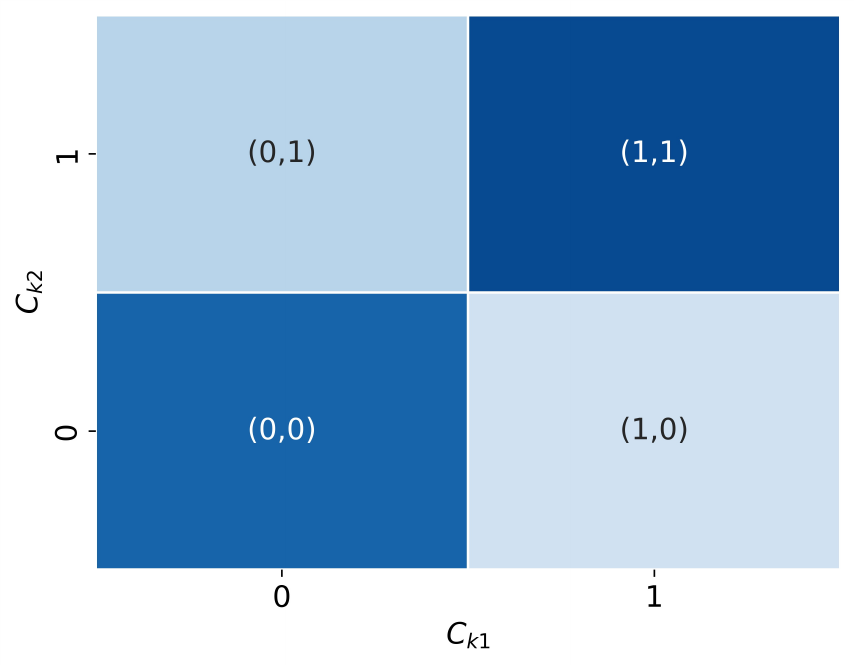}\label{fig:cor_plots_c}}
    \subfloat[$\rho_{C_{k1},C_{k2}}<0$ (invalid)]{\includegraphics[width=0.4\linewidth]{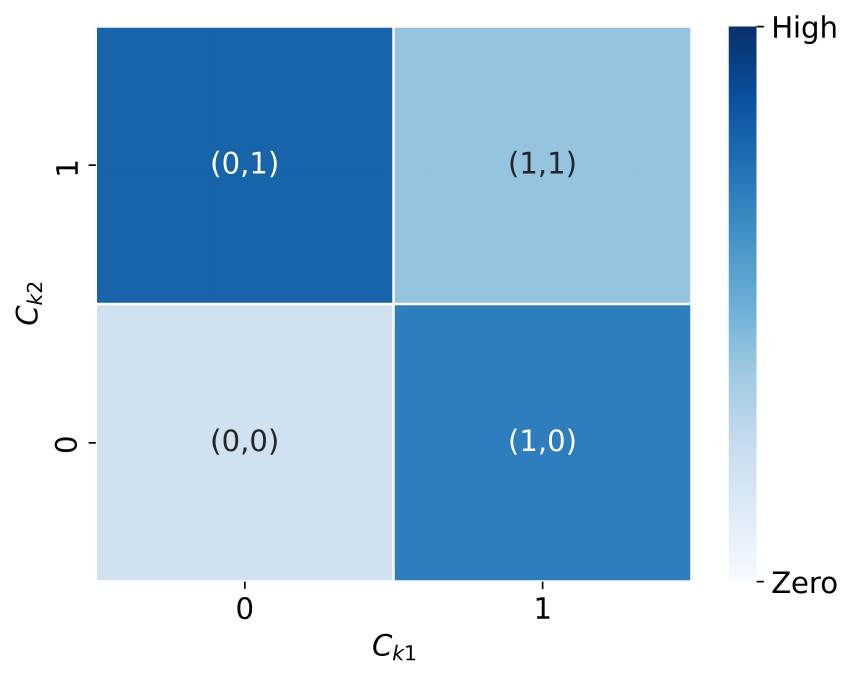}\label{fig:cor_plots_d}}
    \caption{Four cases of Pearson correlation ($\rho$) for ($C_1, C_2$) in multiple trials, which fully characterize the relationship between two Bernoulli random variables. The darkness of the color indicates occurrence frequency. A negative correlation is physically invalid in the setting of a BATS code.}
    \label{fig:cor_plots}
\end{figure}
\JQ{%
The dependence relations represented by a Tanner graph can be exceedingly complicated. To develop some intuition, let us consider a simplified model where an arbitrary variable node is connected to two check nodes. Since we only consider two check nodes in this model, the Pearson correlation coefficient can be used to measure the relation between them.}

Generally, the correlation coefficient is not equivalent to the dependence measured by a more general measure, the mutual information. For example, two random variables being independent implies that the correlation coefficient is zero, but the converse is not always true because the correlation coefficient only measures linear dependencies. 
However, as each node can be represented by a Bernoulli indicator random variable, only linear dependencies can exist between two nodes~\cite{ber}.

Here, we consider a variable node $\tilde{V}$ connected to two check nodes, $\tilde{C_1}, \tilde{C_2}$. 
Consider the Pearson correlation of two sets of data obtained by sampling two random variables, $C_1$ and $ C_2$, respectively, for multiple trials.\footnote{By multiple trials, we mean that decoding is performed multiple times with the same Tanner graph, rank distribution, and other BATS configurations.} 
For each trial, we will obtain a pair of binary indicator numbers. Thus we can plot the pairs in a 2-dimensional space as a heat map to indicate the occurrence frequency for different $(C_1, C_2)$ pairs. The Pearson correlation coefficient $\rho_{C_1, C_2}$ can be calculated to measure the correlation between $C_1$ and $C_2$. \cref{fig:cor_plots} shows four possible cases and the corresponding correlation value range. As discussed, the correlation coefficient is not generally equivalent to the dependence, as the former only measures the linear relationship between random variables while the dependence also measures the nonlinear relationship. 
However, uncorrelatedness and independence are equivalent for multivariate Bernoulli distribution~\cite{ber}, which means that $C_1$ and $C_2$ can only have linear relationships.  
This is further illustrated in \cref{fig:cor_plots}, which lists all possible relationships between two Bernoulli random variables.
When the occurrence of $(0,0)$ and $(1,1)$ dominates, we will have a high positive correlation due to strong positive linear relations between the two random variables as shown in \cref{fig:cor_plots_a} and \cref{fig:cor_plots_c}. When the occurrence of $(0,0)$, $(0,1)$, $(1,0)$ and $(1,1)$ are all the same as shown in \cref{fig:cor_plots_b}, $C_1$ and $C_2$ will have a zero correlation. Note that a negative correlation is meaningless and physically invalid in our context, as one node being undecodable cannot increase the decodability of another node and vice versa. In other words,  $(0,1)$ and $(1,0)$ will never dominate in the trials as shown in \cref{fig:cor_plots_d}. Therefore, without loss of generality, we only consider that $\rho_{C_1, C_2}\in[0,1]$. 

Since a variable node is decodable when at least one of its neighboring check nodes becomes decodable, we can write $V = C_1+C_2-C_1C_2$, where $V, C_1$ and $C_2$ are Bernoulli random variables. Specifically, we have $C_1\sim \textrm{Bernoulli}(\alpha_1)$ and $C_2\sim \textrm{Bernoulli}(\alpha_2)$. Then the expectation of $V$ can be written as
\begin{IEEEeqnarray}{rCl}
\label{eq:exp-v}
    \mathbb{E}(V)&=&\mathbb{E}(C_1)+\mathbb{E}(C_2)-\mathbb{E}(C_1C_2)\IEEEnonumber\\
                   &=&\alpha_1 + \alpha_2 - \mathbb{E}(C_1C_2).
\end{IEEEeqnarray}
The Pearson correlation of $(C_1, C_2)$ is given by 
\begin{equation}
\label{eq:cor}
    \rho_{C_1,C_2}=\frac{\mathbb{E}(C_1C_2)-\alpha_1\alpha_2}{\sqrt{\alpha_1(1-\alpha_1)\alpha_2(1-\alpha_2)}}.
\end{equation}
Substitute $\mathbb{E}(C_1C_2)$ in (\ref{eq:cor}) to (\ref{eq:exp-v}), we can write $\mathbb{E}(V)$ as a function of $\rho_{C_1,C_2}$,
\begin{equation}
\label{eq:exp-v-cor}
    \mathbb{E}(V)=\alpha_1 + \alpha_2 -\alpha_1\alpha_2 - \rho_{C_1,C_2}\sqrt{\alpha_1(1-\alpha_1)\alpha_2(1-\alpha_2)}.
\end{equation}

When $\alpha_1$ and $\alpha_2$ are fixed, \cref{eq:exp-v-cor} suggests that the expectation of $V$ decreases as $\rho_{C_1,C_2}$ increases, and it is minimized when $\rho_{C_1,C_2}=1$. Namely, only $(0,0)$ and $(1,1)$ occur in the trials, which can be achieved by complete dependence as defined in (\ref{eq:complete-dep}). 
It is maximized when $\rho_{C_1,C_2}=0$, that is, when $C_1$ and $C_2$ have no correlation with each other. 
We can see that the complete dependence of check nodes is a sufficient condition to achieve the lower bound of the decoding rate, and the independence of check nodes is a sufficient condition to achieve the upper bound. 
In addition, \cref{eq:exp-v-cor} also characterizes the change of decoding rate along with the correlation coefficient, showing that a higher correlation among check nodes decreases the decoding rate of variable nodes.
However, this correlation analysis assumes the variable node is connected to only two check nodes. 
\JQ{%
Therefore, we will investigate a more general case with conditional probability in the next section.}

\subsection{Dependence Bounds Decodable Probability}
\label{sec:dep-prob}

\begin{figure}
\centering
\includegraphics[width=0.7\linewidth]{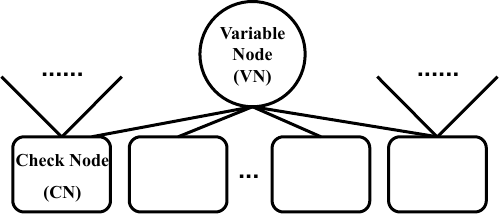}
\caption{A variable node (VN) is connected to multiple check nodes (CNs). All other connections are arbitrary.}
\label{fig:tanner-simp_model}
\end{figure}

Without loss of generality, we consider a random variable node connected to multiple check nodes, and all other connections are arbitrary, as shown in \cref{fig:tanner-simp_model}.
In BATS codes, the decodable probability of a variable node depends on its neighboring check nodes. The variable node will be decodable if at least one of its neighboring check nodes becomes decodable. On the other hand, the decodable probability of a check node depends on how many of its neighboring variables can be decoded during the BP decoding if the check node's rank is insufficient when the BP starts. 



Consider a Tanner graph defined by $\mathcal{T}=(\tilde{C}, \tilde{V}, \tilde{E})$.
Let $\tilde{C}\triangleq\{\tilde{C}_1, \tilde{C}_2, ..., \tilde{C}_N\}$ and $\tilde{V}\triangleq\{\tilde{V}_1, \tilde{V}_2, ..., \tilde{V}_K\}$ be disjoint sets representing the check nodes and variable nodes, respectively.
Set $\tilde{E}$ contains edges that connect the nodes in $\tilde{V}$ and the nodes in $\tilde{C}$. 
Here, we assume that the coding coefficients are chosen independently from the non-zero elements of the base field according to the uniform distribution.
With this setting, we associate each node in $\mathcal{T}$ with a Bernoulli random variable as an indicator for its decodability. Let $C\triangleq\{C_1, C_2, ..., C_N\}$ and $V\triangleq\{V_1, V_2, ..., V_K\}$, where $C_n$ corresponding to $\tilde{C}_n$ and $V_k$ corresponding to $\tilde{V}_k$, $1\leq n\leq N$ and $1\leq k\leq K$, are random Bernoulli variables with parameters of $\alpha_n$ and $\beta_k$, respectively, where $\alpha_n, \beta_k \in (0, 1]$. 
Specifically, we let $P(C_n=1)=\alpha_n$ be the decodable probability of $\tilde{C}_n$ and $P(V_k=1)=\beta_k$ be the decodable probability of $\tilde{V}_k$. Accordingly, we have $P(C_n=0)=1 - \alpha_n$ and $P(V_k=0)=1 - \beta_k$ for the probability of being undecoded. 

To analyze the decodable probability of a variable node and understand how it is affected by its neighboring check nodes,  we first prove an intuitive lemma, which states that when some check nodes are known to be undecodable, the undecodable probability of another check node will either be increased or remain unchanged.

\begin{lemma} \label{lemma:undecoable}
    Assume that a variable node is connected to $n$ check nodes, $\tilde{C}_{1}, \tilde{C}_{2}, ..., \tilde{C}_{n}$, where $n\geq1$. Then for all $i\in [1,n]$ and all $\mathcal{J}\subseteq[1,n]\backslash\{i\}$, 
    \begin{equation}
        P\left(C_{i}=0\right) \leq P\left(C_{i}=0 | C_{j}=0, \ j \in\mathcal{J}\right).\label{eq:undecoable}
    \end{equation}
\end{lemma}
\begin{IEEEproof}
    Let $i\in[1,n]$ be fixed. We will first prove the inequality
    \begin{equation}
        P\left(C_{i}=1\right) \geq P\left(C_{i}=1 |  C_{j}=0, \ j \in\mathcal{J}\right),
        \label{eq:undec_one}
    \end{equation}
    which can be used to prove (\ref{eq:undecoable}) by considering 
    \begin{align}
        &P(C_{i}=1 | C_{j}=0, \ j \in\mathcal{J}) \nonumber\\ 
        &=1-P(C_{i}=0 | C_{j}=0, \ j \in\mathcal{J}), \nonumber
    \end{align}
    and 
    \begin{equation}
        P(C_{i}=1)=1-P(C_{i}=0). \nonumber
    \end{equation}
    We will prove (\ref{eq:undec_one}) by induction on $|\mathcal{J}|$. Firstly, when $|\mathcal{J}|=0$, \ie, $\mathcal{J}=\varnothing$, (\ref{eq:undec_one}) obviously holds. 
    
    Now assume (\ref{eq:undec_one}) holds for all $\mathcal{J}\subseteq[1,n]\backslash\{i\}$, s.t. $|\mathcal{J}|=k$, and we want to prove that it also holds for $|\mathcal{J}|=k+1$, where $0\leq k\leq n-2$. 
    Now consider a subset $\mathcal{J}$ of $[1,n]\backslash\{i\}$ of size $k+1$, and let $\alpha$ be an arbitrary element in $\mathcal{J}$. Then
    \begin{align}
         & P(C_{i}=1 | C_j=0, j\in\mathcal{J})\nonumber \\
         &= \frac{P(C_i=1, C_{\alpha}=0|C_j=0, j\in\mathcal{J}\backslash\{\alpha\})}{P(C_{\alpha}=0|C_j=0, j\in\mathcal{J}\backslash\{\alpha\})} \nonumber \\
         &= \frac{P(C_i=1|C_j=0, j\in\mathcal{J}\backslash\{\alpha\})}{P(C_{\alpha}=0|C_j=0, j\in\mathcal{J}\backslash\{\alpha\})} \nonumber\\
         &\quad - \frac{P(C_i=1, C_{\alpha}=1|C_j=0, j\in\mathcal{J}\backslash\{\alpha\})}{P(C_{\alpha}=0|C_j=0, j\in\mathcal{J}\backslash\{\alpha\})}\nonumber\\
         &= \frac{P(C_i=1|C_j=0, j\in\mathcal{J}\backslash\{\alpha\})}{P(C_{\alpha}=0|C_j=0, j\in\mathcal{J}\backslash\{\alpha\})} \nonumber\\
         &\quad - \frac{P(C_i=1|C_j=0, j\in\mathcal{J}\backslash\{\alpha\})}{P(C_{\alpha}=0|C_j=0, j\in\mathcal{J}\backslash\{\alpha\})}\nonumber\\
         &\quad \times P(C_\alpha=1|C_i=1, C_j=0, j\in\mathcal{J}\backslash\{\alpha\})\nonumber\\
         &\myleq{a} \frac{P(C_i=1|C_j=0, j\in\mathcal{J}\backslash\{\alpha\})}{P(C_{\alpha}=0|C_j=0, j\in\mathcal{J}\backslash\{\alpha\})} \nonumber\\
         &\quad - \frac{P(C_i=1|C_j=0, j\in\mathcal{J}\backslash\{\alpha\})}{P(C_{\alpha}=0|C_j=0, j\in\mathcal{J}\backslash\{\alpha\})}\nonumber\\
         &\quad \times P(C_\alpha=1|C_j=0, j\in\mathcal{J}\backslash\{\alpha\})\nonumber\\
         &= P(C_i=1|C_j=0, j\in\mathcal{J}\backslash\{\alpha\}) \nonumber \\
         &\quad \times \frac{1-P(C_\alpha=1|C_j=0, j\in\mathcal{J}\backslash\{\alpha\})}{P(C_{\alpha}=0|C_j=0, j\in\mathcal{J}\backslash\{\alpha\})} \nonumber\\
         &= P(C_i=1|C_{j}=0, \ j \in\mathcal{J}\backslash\{\alpha\}) \nonumber \\
         &\myleq{b} P(C_i=1), \nonumber
    \end{align}
   where (a) is obtained from the observation that if one check node is decoded, the degree of any other neighboring check node will be decreased by 1, and the rank of that check node will be decreased by at most 1, thus giving
   $P(C_\alpha=1|C_i=1, C_j=0, j\in\mathcal{J}\backslash\{\alpha\}) \geq P(C_\alpha=1| C_j=0, j\in\mathcal{J}\backslash\{\alpha\})$. Since $\alpha$ is an arbitrary element in $\mathcal{J}$, we can obtain (b) from the induction assumption. 
   Thus we have shown that (\ref{eq:undec_one}) holds for all $\mathcal{J}\subseteq[1,n]\backslash\{i\}$, where $|\mathcal{J}|=k+1$. This completes the proof. 
\end{IEEEproof}

With the help of \cref{lemma:undecoable}, we can prove an interesting result that the decodable probability of a variable node is bounded by the decodable probability of its neighboring check nodes. 
Sufficient conditions for achieving the bounds can be expressed in terms of the dependence relations among the check nodes.
Specifically, the upper bound is achieved when all check nodes are mutually independent; the lower bound is achieved when the check nodes are completely dependent.
For $n$ check nodes, we define the \emph{complete dependence} as 
\begin{equation}
 \label{eq:complete-dep}
    P\left(C_1=C_2=...=C_n\right) = 1.
 \end{equation}
We also define the \emph{mutual independence} as 
\begin{equation}
    P(C_1=x_1, ..., C_n=x_n) = P(C_1=x_1)\cdots P(C_n=x_n),
\end{equation}
for all $(x_1, x_2, ..., x_n)\in\{0,1\}^n$.

\begin{theorem}
     For an arbitrary variable node $\tilde{V}$ with $n$ neighboring check nodes, where $n\geq 1$, for $i\in[1,n]$, the decodable probabiliy of $\tilde{V}$ is bounded by \label{theo:edges-depen}
     \begin{equation}
         1-\min_i P(C_{i}=0) \leq P(V = 1) \leq 1-\prod_{k=1}^{n}P(C_k=0). \label{eq:bounds}
     \end{equation}
\end{theorem}

\begin{IEEEproof}
    We see that $\tilde{V}$ is decodable if and only if at least one of its neighboring check nodes becomes decodable. Then
\begin{align}
    &P(V=1) \nonumber\\
    &= 1 - P(C_1=0, C_2=0, ..., C_n=0)\nonumber\\
    &= 1 - P(C_1=0)\prod_{k=2}^{n}P(C_k=0|C_1=0, ..., C_{k-1}=0). \nonumber
\end{align}
    The lower bound in (\ref{eq:bounds}) is obvious as $\prod_{k=2}^{n}P(C_k=0|C_1=0, ..., C_{k-1}=0) \leq 1$, and equality is achieved when all the neighboring check nodes are completely dependent, giving $\prod_{k=2}^{n}P(C_k=0|C_1=0, ..., C_{k-1}=0) = 1$. 
    The upper bound is proved by invoking \cref{lemma:undecoable} as follows: 
    \begin{equation*}
        \begin{cases}
            P(C_2=0)&\leq P(C_2=0|C_1=0) \\
            P(C_3=0)&\leq P(C_3=0|C_1=0,C_2=0) \\
            &\cdots \\
            P(C_n=0)&\leq P(C_n=0 |C_1=0, ..., C_{n-1}=0) \\
        \end{cases}
    \end{equation*}
    \begin{equation*}
        \begin{aligned}
        \Rightarrow &\prod_{k=1}^{n}P(C_k=0) \\ 
            &\leq P(C_1=0)\prod_{k=2}^{n}P(C_k=0|C_1=0, ..., C_{k-1}=0).
        \end{aligned}
    \end{equation*} 
    The upper bound is achievable when the check nodes are mutually independent, which gives $\prod_{k=2}^{n}P(C_k=0|C_1=0, ..., C_{k-1}=0)=\prod_{k=2}^{n}P(C_k=0)$. 
\end{IEEEproof}
\hfill\break
The proof shows that independence and complete dependence are sufficient conditions to achieve the upper bound and the lower bound of the decodable probability, respectively.   

The intuition underlying \cref{theo:edges-depen} is as follows. If the neighboring check nodes are mutually independent, each check node has an independent contribution to the decodability of the variable node, thus achieving the maximum decodable probability. On the other hand, if any of the check nodes are completely dependent, they can be reduced to a single check node, effectively reducing the contribution to the variable node. Thus, we achieve the minimum decodable probability if all check nodes are completely dependent.

\subsection{Implication of Asymptotic Assumption in Dependence}
\begin{figure}
    \centering
    \includegraphics[width=0.9\linewidth]{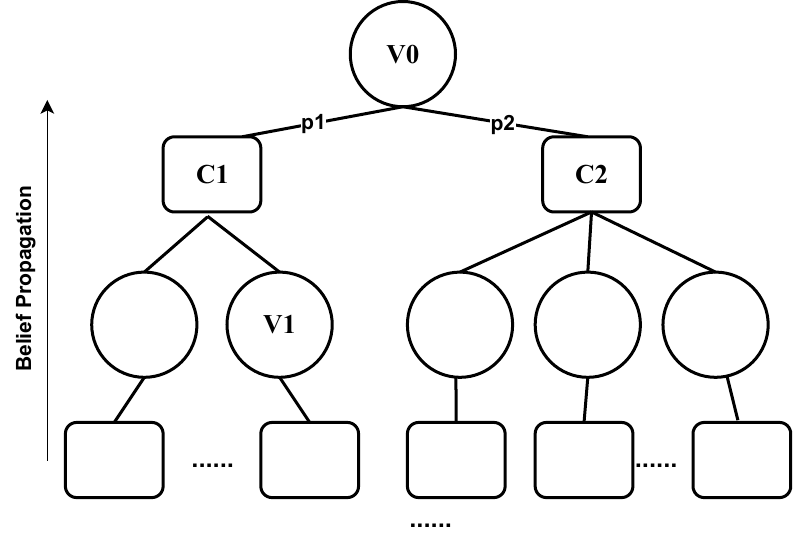}
    \caption{\JQ{Tree representation of a BATS code. The root variable node, V0, is connected to two check nodes, C1 and C2, through connections p1 and p2. Only 4 levels are shown. Belief Propagation decoding is performed level by level from the leaves to the root.}}
    \label{fig:tree}
\end{figure}

\JQ{%
To visualize the physical implication of the asymptotic assumption, we can use a result from the tree analysis~\cite{bats-tree}, which states that when $K$ is sufficiently large, the subgraph expanded from each variable node in the Tanner graph, including all the nodes within its $l$-neighborhood, converges to a tree. Therefore, we can convert a Tanner graph into a tree with $l+1$ levels, where the root is a variable node, as shown in \cref{fig:tree}. According to \cite{bats-tree}, the BP decoding can be applied to the tree level by level from the leaves, and the tree is considered decodable if the root is decodable.} 

\JQ{
In fact, the tree is the least dependent structure when the number of CNs and VNs are fixed. For example, in \cref{fig:tree}, the root V0 has two information propagation paths, p1 and p2, where information that passes through them is independent of each other due to the acyclic nature of a tree. Since the two subtrees of V0 are independent of each other, the upper bound in (\ref{eq:bounds}) is achieved for V0 according to \cref{theo:edges-depen}. On the other hand, if we add an edge between the two subtrees, for example, an edge between V1 and C2, which introduces a cycle, the information that passes through p1 and p2 will be dependent. If enough edges are added between the original two subtrees, causing complete dependence, the lower bound in (\ref{eq:bounds}) will be achieved. Notably, if edges are added between the two subtrees, the tree assumption will also be violated.} 

\JQ{
In finite-length cases, if the BATS code is constructed randomly, there can be many cycles in the Tanner graph, especially when $K$ is moderately small, thus leading to a low decoding rate. However, if we can construct the code with a deterministic structure that reduces the dependence, we can guarantee a better decoding rate than the random BATS.} 

\section{Structured BATS Code}
\label{sec:cs-bats}
Recall that the random BATS code construction relies on sampling a degree distribution to determine the number of edges for each check node and then randomly connecting these edges to the variable nodes. 
Even though the optimized degree distribution ensures the BATS code's asymptotic decodability, the random connection fails to control the structure and the check node dependence, which we have shown to have a significant impact on the decoding rate. 
Additionally, random connection is difficult to implement in hardware. 

Therefore, we propose that \emph{the BATS code should be constructed in a more structured way}. To this end, we design a structured BATS code that is constructed from a base graph using only lightweight cyclic-shift operations, thus called the Cyclic-Shift BATS code (CS-BATS).  
This section shows that the CS-BATS code can achieve better performance and lower computation and implementation complexity with a properly designed base graph. 
Furthermore, the CS-BATS code satisfies the necessary decodability condition and preserves all the desirable properties of the random BATS code, such as the rateless property, the equal protection property, \etc.  

\subsection{Construction of Batches}
Formally, we use the bi-adjacency matrix to represent a Tanner graph, where the columns represent the variable nodes and the rows represent the check nodes. For a Tanner graph $\mathcal{T}$ representing a BATS code of $K$ input symbols, we can write $\mathcal{T}=\{t_0, t_1, ..., t_k, ...\}$, where $t_k$ is the $k$-th row vector in $\mathcal{T}$ and $t_k\in\{0,1\}^{1\times K}$.
\begin{definition}
    (Base Graph)
    The protograph used to construct the code is called the base graph, denoted by $\mathcal{G}=\{g_0, g_1, ..., g_i,... g_{m-1}\}$, where $g_i\in\{0,1\}^{1\times K}, 0\leq i\leq m-1$. 
\end{definition}

\begin{definition}
    (Layer)
    The group of check nodes generated from the base graph with the same number of cyclic shifts is called a layer. Specifically, if a check node is generated with $n$ cyclic shifts, we say it belongs to the $n$-th layer. The check nodes of the base graph form the $0$-th layer. 
\end{definition}

\begin{algorithm}
    \caption{Cyclic-Shift BATS Code Encoding}\label{alg:cs-bats-enc}
    \begin{algorithmic}[1]
        \State \textbf{Input} $\mathcal{G}=\{g_0, ..., g_{m-1}\}$
        \State \textbf{Output} $\mathcal{T}=\{t_0, ..., t_{N-1}\}$ for $N$ batches
        \For{$n$ in $\{0, 1, ..., N-1\}$ }
            \State $t_i = g_{(i\bmod{m})} \ggg \lfloor{i / m}\rfloor$
        \EndFor
    \end{algorithmic}
\end{algorithm}

\begin{figure}
\centering
\includegraphics[width=0.8\linewidth]{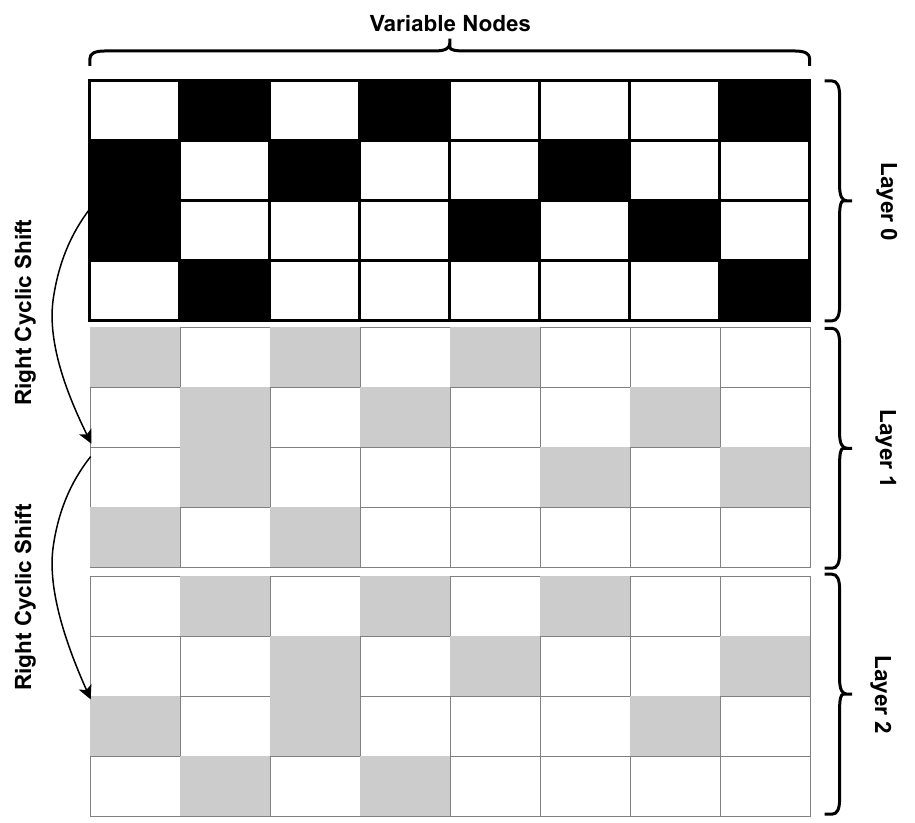}
\caption{\textbf{Cyclic-Shift BATS Code constructed from a Base Graph}. The underlying bipartite graph is represented by its bi-adjacency matrix, where the darkened blocks represent 1s (0 otherwise). The $n$-th layer of the check nodes is generated by right-cyclic-shifting the rows in the base graph $n$ times, where $n\geq 0$ and layer 0 is the base graph. \JQ{Note that the bi-adjacency matrix only represents the connections and does not represent the edge weights in the Tanner graph. The edge weights are represented by the generator matrix $\bm{G}_i\in\mathbb{F}_q^{dg\times M}$}}
\label{fig:cs-bats-enc}
\end{figure}

\cref{alg:cs-bats-enc} describes the proposed procedure to construct a BATS code of $N$ batches from a base graph. Check nodes are constructed from the rows in the base graph using right cyclic-shift operations (denoted by ``$\ggg$''). Notice that the number of variable nodes is the same in $\mathcal{G}$ and $\mathcal{T}$, but $\mathcal{G}$ has much fewer check nodes than $\mathcal{T}$. This procedure is illustrated in \cref{fig:cs-bats-enc}.

\subsection{Bounded Complexity}
\label{sec:bound_comp}
In hardware implementations, data originally stored in a large memory is usually moved in small pieces to a faster buffering memory like on-chip registers for further computations. In the context of the BATS code, input symbols are selected and moved to the buffer for linear combinations.   
As illustrated in \cref{fig:circuit_complexity}a, when the input symbol selection is random and determined on the fly, a large buffer is needed to accommodate the maximum possible degree, thus giving a \textbf{buffering complexity} of $\mathcal{O}(K)$, where $K$ is the total number of input symbols. Additionally, for each position of the buffer, any inputs can be selected and routed to this position, which leads to a \textbf{routing complexity} of $\mathcal{O}(K^2)$.

On the other hand, 
if the input symbol selection is structured like the CS-BATS code, we only need to allocate a buffer with an appropriate size according to the row degrees of the base graph. Therefore, the \textbf{buffering complexity} is $\mathcal{O}(d)$, where $d$ is the maximum degree of the base graph.
Additionally, as the position of selected input symbols is pre-determined, we can just move input symbols to the buffer from a fixed position and cyclic-shift all input symbols in the memory. This gives a \textbf{routing complexity} of $\mathcal{O}(d)$. 

\JQ{%
For example, for a random BATS code, when $K=256$, the maximum degree is also $256$, which means that up to 256 packets could be moved from the data storage to the computation buffer. In hardware implementation, data storage usually consists of a large off-chip memory connected to the computation circuits through interconnections. The computation buffers are usually fast on-chip registers that make parallel processing of data possible. Therefore, we need enough buffers to accommodate the maximum number of packets, which is 256 in this example, even though the maximum degree is sampled with a small probability. If each packet consists of $256$ symbols from $GF(2^8)$, $64$ KBytes of buffers are required. 
In comparison, the maximum degree of a CS-BATS code is determined by the base graph, which is typically much smaller. For example, in \cref{sec:exp}, our experiments use a base graph with a maximum degree of $27$, which requires only $6.75$ KBytes of buffers on hardware.}

\JQ{%
Since packets are randomly selected in the random BATS, every packet could be selected and routed to every register buffer through a multiplexer as shown in \cref{fig:circuit_complexity}, which is a fully connected circuit with $256\times256$ wires that connect the data storage to the buffers. However, if the structure is deterministic, as for the CS-BATS, the packets can be easily routed to the buffer with cyclic shift operations, requiring only $27$ wires in our example.}

\JQ{%
In FPGA, off-chip data is usually accessed by addressing a Direct Memory Access (DMA) Unit. Random data access caused by the random BATS gives the worst-case performance, while the inherent structured access patterns in the CS-BATS can achieve the highest throughput through bursting, coalescing, and other optimization techniques~\cite{hbm}.}

As we can see, the implementation complexity of CS-BATS, determined by the base graph, is significantly reduced compared with the random BATS, especially when $d\ll K$. 
Therefore, with the CS-BATS code, the tradeoff between the complexity and the code performance can be controlled by the design of the base graph. 
It does not rely on reducing $K$ to reduce the complexity, which is important for adapting the implementation to different devices without affecting the code performance. 

\subsection{Preserved Properties}
According to \cref{alg:cs-bats-enc}, an unlimited number of batches can be generated with light-weight cyclic-shift operations, which preserves the rateless property of the BATS code. This is important for the adaptability of the BATS code for achieving reliable communication in a network with unknown or changing channel conditions. 

The random connection in the traditional BATS code ensures that each input symbol has the same probability of being selected by a check node. The CS-BATS code also preserves this property. 
The empirical probability of each variable node being selected converges to the same value as the number of check nodes goes to infinity because the selection position is shifted cyclically on all input symbols. 

\subsection{Base Graph Design}
\label{sec:design}
Based on our previous analysis, this section proposes two empirical conditions for a good base graph. 

\textbf{Convergence to an optimized degree distribution}. 
The row degrees of the base graph should be chosen according to the optimized degree distribution. 
Specifically, we first take the normalized probability, $\Psi'$, of the $m$ largest probability masses of the optimized degree distribution $\Psi$, which is obtained by minimizing $\theta$ in (\ref{eq:bats-opt-obj}) such that the conditions in (\ref{eq:bats-opt-nec}) and (\ref{eq:bats-opt-sum}) are satisfied.  Assume that $\Psi$ is in descending order of the probability masses. Then
\begin{equation}
    \Psi'_d = \frac{\Psi_d}{\sum^m_{j=1}\Psi_{j}}.
\end{equation}
Note that when $m\rightarrow\infty$, $\Psi'_d\rightarrow\Psi_d$.
Let us now consider designing the base graph $\mathcal{G}$ with an aim to mimic $\Psi'_d$. 
Let
\begin{equation}
    \gamma_d=\#\{g\in\mathcal{G}| \textrm{deg}(g) = d\},
\end{equation}
where $\textrm{deg}(g)$ is the degree of row $g$. Then in order to mimic $\Psi'_d$, for $1\leq d \leq m$, $\gamma_d$ should be chosen such that 
\begin{equation}
\label{eq:conver_opt_deg}
    \begin{cases}
        \sum^K_{d=1}\gamma_d = m \\
        \gamma_d = \lceil m\Psi'_d\rceil \ \textrm{or}\ \lfloor m\Psi'_d\rfloor.
    \end{cases}
\end{equation}

Condition (\ref{eq:conver_opt_deg}) suggests that the degrees of the base graph should be designed to match the largest $m$ probability masses in the optimized degree distribution. This ensures the necessary condition (\ref{eq:bats-opt-nec}) for the decodability of the BATS code to be satisfied. 


\begin{figure*}[t]
    \centering
    \subfloat[Inactivation Decoding]{\includegraphics[width=0.4\linewidth]{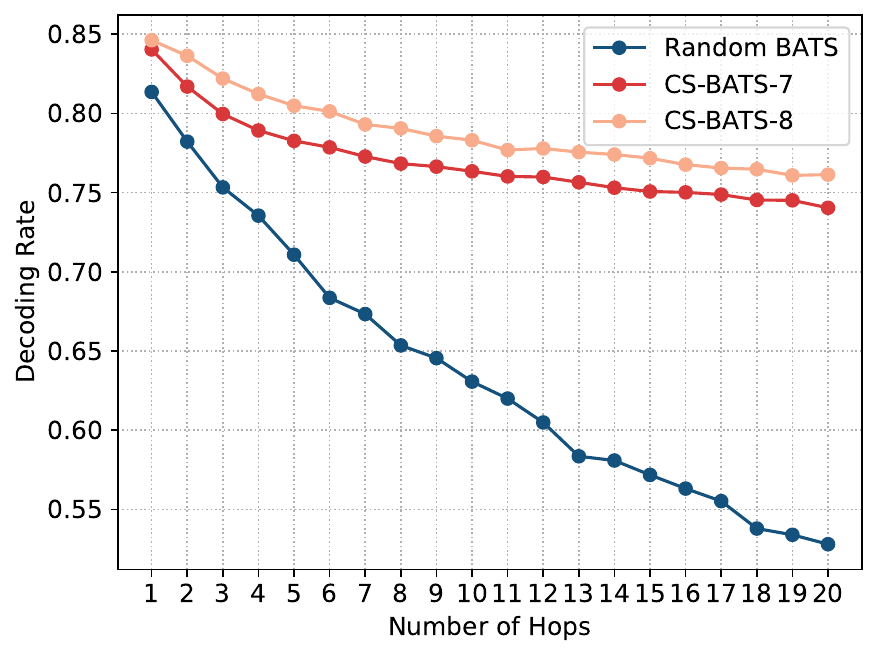}\label{fig:decoding_rates1a}}
    \subfloat[BP Decoding]{\includegraphics[width=0.4\linewidth]{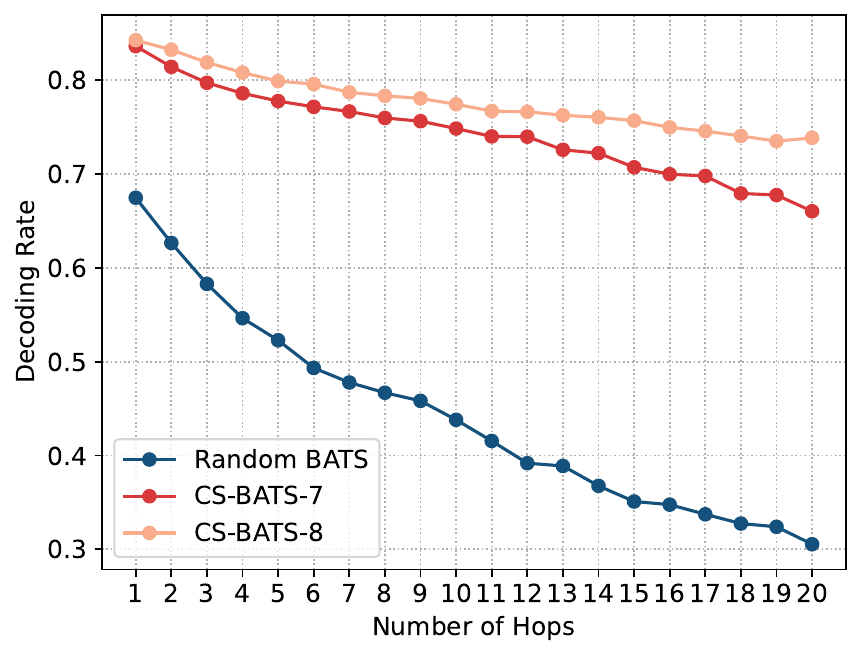}\label{fig:decoding_rates1b}}
    
    \caption{\textbf{Experiment 1.} Decoding rates vs the number of hops plots under (a) inactivation decoding and (b) BP decoding. Average results from 2000 randomly generated instances are reported for the Random BATS. Average results from 500 repetitions using the designed base graph are reported for the CS-BATS. \JQ{CS-BATS-7: CS-BATS code using a base graph with 7 rows; CS-BATS-8: CS-BATS code using a base graph with 8 rows.}}
    \label{fig:decoding_rates1}
\end{figure*}

\begin{figure*}[t]
    \centering
    \subfloat[Inactivation Decoding]{\includegraphics[width=0.4\linewidth]{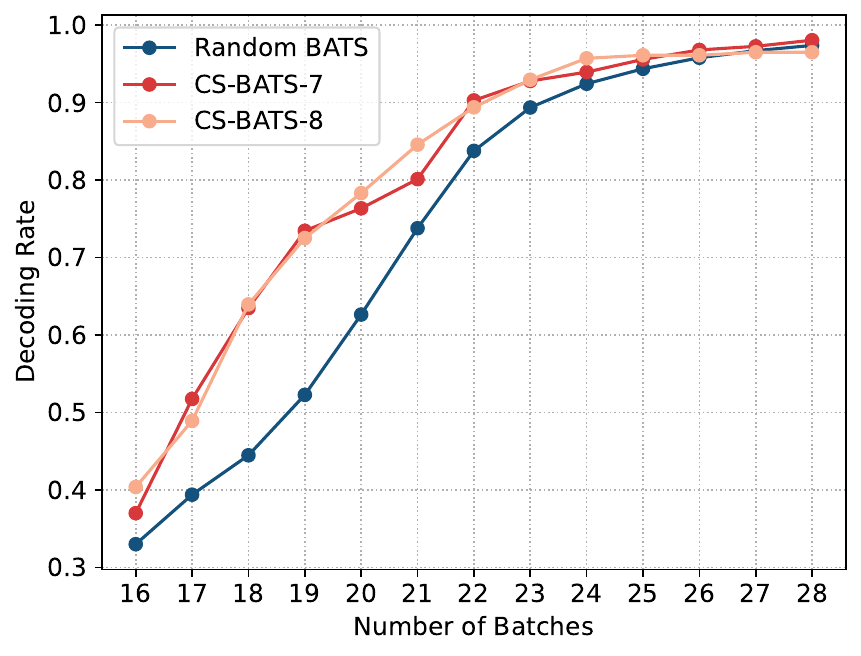}\label{fig:decoding_rates2a}}
    \subfloat[BP Decoding]{\includegraphics[width=0.4\linewidth]{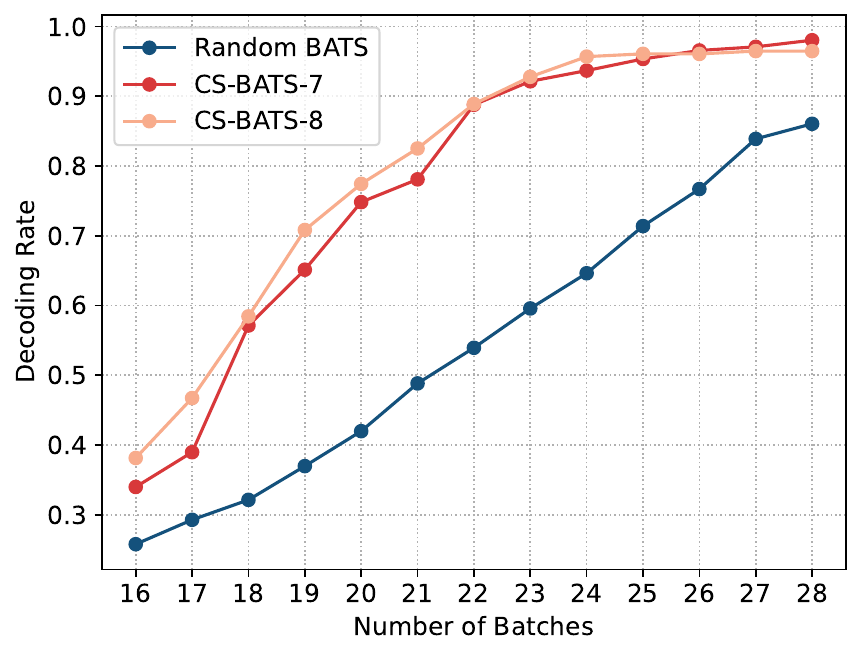}\label{fig:decoding_rates2b}}
    \caption{\textbf{Experiment 2.} Decoding rates vs the number of batches under (a) inactivation decoding and (b) BP decoding. \JQ{CS-BATS-7: CS-BATS code using a base graph with 7 rows; CS-BATS-8: CS-BATS code using a base graph with 8 rows.}}
    \label{fig:decoding_rates2}
\end{figure*}

\begin{table*}[t]
\caption{\textbf{Numerical Results}. Average results from 2000 instances of randomly generated graphs are reported for the Random BATS. Each experiment is repeated 10 times. Average results from 500 repetitions using the designed base graph are reported for the CS-BATS. Higher decoding rates are highlighted.}
    \centering
    \resizebox{\textwidth}{!}{\begin{tabular}{|c|c|c|ccccc|ccccc|}
         \hline
         \multicolumn{3}{|c|}{\multirow{2}{*}{}} & \multicolumn{5}{c|}{Number of Batches (code rate) (10 hops)} & \multicolumn{5}{c|}{Number of Hops (20 batches)} \\
         \cline{4-13}
         
         \multicolumn{3}{|c|}{} & 16 (1.0) & 18 (1.125) & 20 (1.25) & 22 (1.375) & 24 (1.5) & 1 & 5 & 10 & 15 & 20 \\
         \hline
         
         \multirow{4}{*}{\shortstack{Decoding\\Rate}}&\multirow{2}{*}{\shortstack{Inac}} & 
         CS  & \textbf{0.37}\tiny{$\pm0.09$}  & \textbf{0.63}\tiny{$\pm0.07$} & \textbf{0.76}\tiny{$\pm0.03$} & \textbf{0.90}\tiny{$\pm0.06$} & \textbf{0.94}\tiny{$\pm0.03$} & 
         
         \textbf{0.84}\tiny{$\pm0.04$} & \textbf{0.78}\tiny{$\pm0.03$} & \textbf{0.76}\tiny{$\pm0.03$} & \textbf{0.75}\tiny{$\pm0.03$} & \textbf{0.74}\tiny{$\pm0.04$} \\

         & & Rand & 0.33\tiny{$\pm0.17$} & 0.44\tiny{$\pm0.20$} & 0.63\tiny{$\pm0.19$} & 0.84\tiny{$\pm0.12$} & 0.92\tiny{$\pm0.07$} & 
         
         0.82\tiny{$\pm0.11$} & 0.71\tiny{$\pm0.14$} & 0.63\tiny{$\pm0.19$} & 0.57\tiny{$\pm0.20$} & 0.52\tiny{$\pm0.21$}\\\cline{2-3}
         
         & \multirow{2}{*}{BP} & 
         CS  & \textbf{0.35}\tiny{$\pm0.08$} & \textbf{0.57}\tiny{$\pm0.12$} & \textbf{0.75}\tiny{$\pm0.06$}& \textbf{0.89}\tiny{$\pm0.06$} & \textbf{0.93}\tiny{$\pm0.03$} & 
         
         \textbf{0.84}\tiny{$\pm0.04$} & \textbf{0.78}\tiny{$\pm0.04$} & \textbf{0.75}\tiny{$\pm0.06$} & \textbf{0.71}\tiny{$\pm0.12$} & \textbf{0.67}\tiny{$\pm0.015$}\\

         & & Rand & 0.26\tiny{$\pm0.15$} & 0.32\tiny{$\pm0.18$} & 0.42\tiny{$\pm0.20$} & 0.54\tiny{$\pm0.22$} & 0.65\tiny{$\pm0.21$} & 
         
         0.66\tiny{$\pm0.15$} & 0.53\tiny{$\pm0.18$} & 0.42\tiny{$\pm0.20$} & 0.37\tiny{$\pm0.20$} & 0.31\tiny{$\pm0.19$} \\
         \hline
         
         \multicolumn{2}{|c|}{\multirow{2}{*}{$\#$Inact Symbols}} & 
         CS  & 17\tiny{$\pm2.8$} & 11\tiny{$\pm2.0$} & 4.4\tiny{$\pm1.4$} & 0.4\tiny{$\pm0.7$} & 0.4\tiny{$\pm0.7$} & 
         
         1\tiny{$\pm1.2$} & 3.3\tiny{$\pm1.5$} & 4.4\tiny{$\pm1.4$} & 5.4\tiny{$\pm1.5$} & 5.8\tiny{$\pm1.6$}\\
         
         \multicolumn{2}{|c|}{} & 
         Rand & 43\tiny{$\pm21$} &30\tiny{$\pm16$} & 18\tiny{$\pm13$} & 8\tiny{$\pm10$} & 4\tiny{$\pm8$} & 
         
         10.5\tiny{$\pm13$} & 16.4\tiny{$\pm13$} &  17.9\tiny{$\pm12$} & 20.7\tiny{$\pm12$} & 23.2\tiny{$\pm14$}   \\
         \hline
         
         \multicolumn{2}{|c|}{\multirow{2}{*}{$\#$Edges}} & 
         CS  & 257 & 285 & 324 & 362 & 388 
          & \multicolumn{5}{c|}{\multirow{2}{*}{\diagbox[innerwidth=25em, height=2\line]{}{} }} \\
         \multicolumn{2}{|c|}{} & 
         Rand & 742\tiny{$\pm128$} & 818\tiny{$\pm157$} & 890\tiny{$\pm165$} & 940\tiny{$\pm162$} & 1018\tiny{$\pm183$} & & & & & \\
         \hline
    \end{tabular}}
    \label{tab:results}
\end{table*}

\begin{table*}[t]
    \caption{\textbf{The column degree design outperforms the random base graph.} CS: randomly generated base graph; CS-CD: random base graphs with the column degree design. For each algorithm, 500 base graphs are randomly generated with the same degrees, and the average decoding rate using BP decoding is reported.}\label{tab:degree_matching}
    \centering
    \resizebox{\textwidth}{!}{\begin{tabular}{|c|c|cccccccccc|}
         \hline
         \multicolumn{2}{|c|}{\multirow{2}{*}{}} & \multicolumn{10}{c|}{Number of Batches (code rate) (10 hops)} \\
         \cline{3-12}
         
         \multicolumn{2}{|c|}{} & 19  & 20  & 21  & 22  & 23  & 24  & 25  & 26  & 27  & 28 \\
         \hline
         
         \multirow{2}{*}{\shortstack{Decoding Rate}} & CS & 0.70\tiny{$\pm0.03$} & 0.73\tiny{$\pm0.03$} & 0.76\tiny{$\pm0.03$} & 0.77\tiny{$\pm0.03$} & 0.78\tiny{$\pm0.03$} & 0.79\tiny{$\pm0.03$} & 0.80\tiny{$\pm0.03$} & 0.83\tiny{$\pm0.03$} & 0.83\tiny{$\pm0.03$} & 0.85\tiny{$\pm0.03$}\\
         
         & CS-CD  & \textbf{0.74}\tiny{$\pm0.04$}  & \textbf{0.79}\tiny{$\pm0.02$} & \textbf{0.83}\tiny{$\pm0.02$} & \textbf{0.84}\tiny{$\pm0.02$} & \textbf{0.85}\tiny{$\pm0.02$} & \textbf{0.86}\tiny{$\pm0.02$} & \textbf{0.88}\tiny{$\pm0.02$} & \textbf{0.89}\tiny{$\pm0.01$} & \textbf{0.90}\tiny{$\pm0.02$} & \textbf{0.91}\tiny{$\pm0.02$}\\
         \hline
    \end{tabular}}
\end{table*}

\textbf{Column degree design}.
After the row degrees are chosen, we need to connect the check nodes and variable nodes in the base graph accordingly, which, in effect, determines the number of neighboring check nodes for each variable node in the base graph. As the variable nodes are represented by the columns in the bi-adjacency matrix, we also refer to the number of neighboring check nodes for a variable node as the column degree. 

In an ideal case, mutual independence among check nodes is desired to achieve the decoding rate upper bound. 
Even though this condition is usually too strong for building the complete Tanner graph, we can still obtain a graph with a ``less dependent'' structure by minimizing the dependence of check nodes in the base graph. 
Specifically, we will choose columns with the smallest column degrees for a given row degree, as shown in \cref{alg:degree_matching}, which leads to a base graph with balanced variable node connections.
Notice that when there are multiple columns with the smallest column degree, a column will be picked randomly.  
\JQ{A balanced variable node connectivity is needed to avoid large dependence between check nodes when the check node degrees are fixed.}

\begin{algorithm}[h]
    \caption{Column degree design}\label{alg:degree_matching}
    \begin{algorithmic}[1]
        \State \textbf{Input} Pre-determined degrees $\{d_0, d_1,..., d_{m-1}\}$
        \State \textbf{Output} Base graph $\mathcal{G}=\{g_0, g_1,... , g_{m-1}\}$
        \State // ensure variable node coverage
        \State variable\_idx = $[0, 1,... , K-1]$ 
        \For{$i$ in $[0, 1,..., m-1]$}
            \State idx $\leftarrow$ the smallest $d_i$ indices from variable\_idx with the smallest column degrees.  
            \State $g_i = 0$
            \State $g_i[\textrm{idx}] = 1$
        \EndFor
    \end{algorithmic}
\end{algorithm}

\subsection{Experiments}
\label{sec:exp}
In this section, we will compare the CS-BATS code with the random BATS code in terms of performance, the number of edges, and the number of inactivation symbols. 
In $GF(2^8)$, we use a packet number $K$ of $256$, a packet size $pk$ of $256$, and a batch size $M$ of $16$ for the BATS code. A $10\%$ i.i.d. packet loss for the links is applied during the simulation.
Randomly generated generator matrices from $GF(2^8)$ will be used for all experiments.
The optimized degree distributions $\Psi_{BP}$ and $\Psi_{inact}$ are obtained for BP decoding and inactivation decoding, respectively.
For the random BATS code, we generated 2000 random instances of the code for each experiment and report the average results and standard deviation. Each instance was repeated 10 times in the simulation for reliable results.  

For CS-BATS, we choose a base graph of dimension $7\times 256$ (\ie, $m=7$, $K=256$) generated by \cref{alg:degree_matching}. Starting from the optimized degree distributions $\Psi_{BP}$, we applied the method described in \cref{sec:design} to obtain the following degree design for the base graph: 
\begin{equation*}
    \{11, 12, 14, 14, 19, 20, 27\},
\end{equation*}
which will be used in \cref{alg:degree_matching}.
\JQ{Notice that the number of rows in the base graph ($m$) determines the size of the base graph. The implementation complexity increases as $m$ increases, approaching the complexity of the random BATS. Additionally, the design space of the CS-BATS increases as $m$ increases, thus potentially leading to the design of a better code. In the extreme case that $m$ is very large, we can pre-define the structure of every generated batch so that they are mutually independent. In practice, $m$ should be determined according to the implementation requirement.}
\JQ{To demonstrate the impact of $m$ in the simulations, we generate a base graph with $m=8$ with the same procedure. The following row degrees are used:}
\begin{equation*}
    \{11, 12, 14, 14, 16, 19, 20, 27\}.
\end{equation*}

\JQ{We will also show the effectiveness of \cref{alg:degree_matching} in designing a good base graph by comparing the average decoding rate of 500 instances of randomly generated base graphs and the base graphs generated by \cref{alg:degree_matching}.}

\textbf{Decoding Rates}.
Recall that the decoding rate is defined as the number of decoded symbols divided by the number of input symbols, which is the most direct metric to measure the decoding effectiveness of a code under the erasure channel model. In practice, a fixed-rate precode can be added to ensure full recovery provided that the BATS code achieves a certain decoding rate~\cite{bats-tit}. However, we consider the BATS code without precoding here. 

Two experiments were designed to compare the performance of the CS-BATS code and the random BATS code from different perspectives. In the first experiment, we fix the number of batches to 20 ($\textrm{code rate}=MN/K=1.25$) and vary the number of hops. This experiment shows how the performance changes as the number of hops increases, which is of great practical interest for a BATS code.  
The second experiment fixes the number of hops to 10 and varies the number of batches, which investigates the change in performance as the number of batches $N$ changes. Specifically, we start from $N=16$ ($\textrm{code rate}=1.0$), and end at $N=28$ ($\textrm{code rate}=1.75$).  

In \cref{fig:decoding_rates1} and \cref{fig:decoding_rates2}, we plot the results for inactivation decoding and BP decoding, respectively. 
Part of the numeric results is also shown in \cref{tab:results}.
From \cref{fig:decoding_rates1}, we can see that the CS-BATS code retains a higher and more stable decoding rate for both inactivation decoding ($0.84$ to $0.74$) and BP ($0.83$ to $0.66$) as the number of hops increases from 1 to 20. In contrast, the decoding rate of the random BATS code degrades from $0.82$ to $0.51$ for inactivation decoding and from $0.66$ to $0.31$ for BP decoding as the hop increases. 
From \cref{fig:decoding_rates2}, as the number of batches increases, the decoding rate of the CS-BATS also outperforms the decoding rate of the random BATS, especially when BP decoding is used. Even though the performance is close in the higher decoding rate region when inactivation decoding is used, fewer inactivation symbols are used by CS-BATS, leading to a smaller decoding complexity as shown in \cref{tab:results}.

By comparing \cref{fig:decoding_rates1a} and \cref{fig:decoding_rates1b}, and comparing \cref{fig:decoding_rates2a} and \cref{fig:decoding_rates2b}, we see that a significant performance gap exists between inactivation decoding and BP decoding for the random BATS. Thus, in practice, the random BATS code is usually decoded with inactivation decoding. However, this gap is greatly reduced by the CS-BATS code, making BP decoding a feasible choice for some resource-limited applications. 

\JQ{When comparing the performance of $m=7$ and $m=8$ for the CS-BATS, we observe that the latter outperforms the former in general, which confirms our analysis that a larger base graph offers more design choices. In order to make a fair comparison between $m=7$ and $m=8$, we search among the base graphs generated using \cref{alg:degree_matching} for the ones that contain the most balanced variable node connections after constructing the Tanner graph. } 

\textbf{Number of Edges}.
In \cref{tab:results}, we also record the number of 
edges of the Tanner graph with different numbers of check nodes (batches), as the total number of edges is usually related to the encoding and decoding complexity. The average results are also reported across multiple instances of Tanner graphs. Generally, we can see that the number of edges increases with the number of batches. Graphs constructed with CS-BATS code have much fewer edges (around 30\% of the random BATS), which are controlled by the degrees of the base graph. 

\textbf{Number of Inactivation Symbols}.
The number of inactivation symbols represents the computation needed in inactivation decoding~\cite{bats-tit}. Decoding processes with fewer inactivation symbols have less computation, leading to higher throughput and shorter delay. Therefore, we also compare the number of inactivation symbols used when decoding the graphs constructed by the CS-BATS with those constructed by the random BATS. As shown in \cref{tab:results}, the CS-BATS uses much fewer inactivation symbols than the random BATS, indicating a significant increase in computational efficiency. 

\textbf{Column Degree Design}. 
In \cref{tab:degree_matching}, we compare the results obtained from randomly generated base graphs and those generated by \cref{alg:degree_matching}. Specifically, 500 instances of randomly generated base graphs and another 500 randomly generated base graphs with column degree designing were simulated with different numbers of batches. We observe a higher decoding rate for base graphs with column degree designing in \cref{tab:degree_matching}.

\begin{algorithm}[t]
    \caption{Layered Decoding}\label{alg:layer-dec}
    \begin{algorithmic}[1]
        \State \textbf{Input} Received Tanner graph $\mathcal{T}$ with $N$ batches
        \State \textbf{Output} Decoded packets $\{v_0, v_1,...\}$
        \For{layer $\mathcal{T}_k$ in $\{\mathcal{T}_0,\mathcal{T}_1,..., \mathcal{T}_{\lceil\frac{N}{m}\rceil} \}$}
            \For{$c$ in the set of CNs in $\mathcal{T}_k$} \label{bp}
                \State $\Call{BP}{c}$
            \EndFor
            \For{$v$ in the set of undecoded VNs in $\mathcal{T}_k$}
                \State Inactivate $v$
                \For{$c_v$ in the set of neighboring CNs of $v$}
                    \State $\Call{BP}{c_v}$
                \EndFor
            \EndFor
            \If{linear constraint rank = $\#$inactivated symbols}
                \State Solve inactivation symbols
                \State Substitute into involved VNs
            \EndIf
        \EndFor
        \State
        \Procedure{BP}{$c$}
             \If{rank($c$) $=$ deg($c$) and $c$ is unsolved}
                    \State Solve $c \rightarrow \{v_{k1}, v_{k2}, ...\}$
                    \State Collect linear constraints for inactivated symbols
                    \For{$c_k$ in the set of neighboring CNs $\{c_{k1}, c_{k2}, ...\}$}
                        \State Substitute $v_{k}$ into $c_k$
                        \State $\Call{BP}{c_k}$ \Comment{Call BP again}
                    \EndFor
            \EndIf
        \EndProcedure
    \end{algorithmic}
\end{algorithm}

\section{Layered Decoding}
\label{sec:layer-dec}
As a further demonstration of how the CS-BATS reduces the hardware implementation complexity, we will propose a complexity-reduced and flexible decoding algorithm tailored for hardware.  

The number of batches to be transmitted in a wireless communication session depends on the channel condition during that session. In particular, more batches are needed if the channel condition is poor. 
Despite the uncertainty on the number of arrived batches, an intuitive decoder implementation is to set a maximum processing ability, allocating enough hardware resources in advance, such as memory buffers for storing received batches and computation circuits for solving and substituting symbols. 
However, this limits the flexibility and decoding ability of the design. 
Ideally, the decoder should be able to accommodate a potentially unlimited number of batches, and the hardware consumption should be flexible and scalable. 
The decoding algorithm for CS-BATS to be discussed next addresses this issue. 

The Tanner graph constructed by the CS-BATS can be naturally divided into layers according to the number of cyclic shifts. Motivated by this layered structure and inspired by a similar complexity-reducing decoding scheme in LDPC~\cite{ldpc-LD, ldpc-LD-fpga, ldpc-LD-fpga2}, we propose a layered decoding algorithm for the CS-BATS whose implementation complexity is bounded by the size of the base graph. 

Consider \cref{alg:layer-dec} for layered decoding for a base graph with $m$ rows.
In this algorithm, BP and inactivation decoding is performed layer by layer. After the inactivation decoding consumes all the batches in a layer, the received packets of this layer can be discarded, and the decoding process will proceed to decode the next layer. 
Since each layer is decoded sequentially with the same procedure, the same hardware can be used for each layer. 
During the transition from one layer to another, the only growth in memory consumption is due to the linear constraints collected during the BP decoding from the previous layer. 
Generally, the total number of linear constraints that need to be stored depends on the number of inactivation symbols, which is bounded by $\mathcal{O}(K)$.
The previous section shows that the CS-BATS uses only a few inactivation symbols. Thus, the memory consumption caused by the linear constraints is mild. 

Notice that \cref{alg:layer-dec} can also be used by the random BATS code. However, the CS-BATS with embedded layer structure can fully exploit this algorithm with hardware reuse and low implementation complexity. 

As discussed in \cref{sec:bound_comp}, the encoding complexity of the CS-BATS depends only on the base graph. The same applies to the complexity of layered decoding. Thus, the implementation complexity of decoding can also be controlled by the design of the base graph. This leads to higher flexibility and better scalability in hardware design, which is essential for adapting the implementation to various grades of devices.

\section{Discussion and Conclusion}
\label{sec:conclusion}
This paper analyzes the influence of check node dependence on the decoding rate of a BATS code. 
We show that the check node dependence degrades the performance of a BATS code, especially when the number of input symbols is finite.  
Based on the analysis, we propose constructing the code in a more structured way instead of relying on random connections. As an example, a structured BATS code called the Cyclic-Shift BATS is presented, which controls the check node dependence and introduces a new design dimension. 
Conditions supported by empirical experiments are given for designing a well-structured base graph. 

We further demonstrate that the CS-BATS code achieves a significantly higher decoding rate and more stable performance with a smaller and controllable complexity compared with the random BATS. 
Furthermore, we propose a layered decoding algorithm that exploits the layered structure of the CS-BATS code. The implementation complexity of this algorithm is bounded by the size of the base graph. 

The goal of this paper is to present a novel class of BATS codes, the structured BATS, and demonstrate that its performance can be superior to the random BATS when its structure is well-designed. The CS-BATS is a possible instance of this class of codes, which is designed especially for efficient hardware implementations. Further research is necessary to design different variants of the structure BATS tailored for different purposes. 
\JQ{For example, storing the base graph when the code length is large may incur significant overhead. In that case, more flexible construction methods could be used to construct the base graph by lifting a smaller graph.}

\section{Acknowledgement}
The authors would like to thank Prof.~Philip~H.~W.~Leong, Prof. Kin~Hong~Lee, Hoover~H.~F.~Yin, Yulin~Chen, and Fangwei~Ye for the useful discussions and valuable suggestions on this work.

\bibliographystyle{IEEEtran}
\bibliography{IEEEabrv, paper.bib}

\end{document}